\documentclass{emulateapj}
\usepackage{apjfonts}
\usepackage{amsmath}

\def\ergs{\hbox{~ergs~s$^{-1}$}}
\def\kms{\hbox{~km~s$^{-1}$}}
\def\cc{\hbox{~cm$^{-3}$}}

\begin{document}

\title{Ultra-Luminous X-ray Sources: Evidence for Very Efficient
Formation of
Population III Stars Contributing to the Cosmic Near-Infrared
Background Excess?}
\author{Haruka Mii and Tomonori Totani}
\affil{Department of Astronomy, Kyoto University \\
Sakyo-ku, Kyoto 606-8502, Japan}
\email{mii@kusastro.kyoto-u.ac.jp}

\begin{abstract}
Accumulating evidence indicates that some of ultra-luminous X-ray sources
(ULXs) are intermediate mass black holes (IMBHs), but the formation
process of IMBHs is unknown. One possibility is that they were formed
as remnants of population III (Pop III) stars, but it has been thought
that the probability of being an ULX is too low for IMBHs distributed in
galactic haloes to account for the observed number of ULXs. Here we
show that the number of ULXs can be explained by such halo IMBHs passing
through a dense molecular cloud, if Pop III star formation is very
efficient as recently suggested by the excess of the cosmic near-infrared
background radiation that cannot be accounted for by normal galaxy
populations. We calculate the luminosity function of X-ray sources in
our scenario and find that it is consistent with observed data. Our
scenario can explain that ULXs are preferentially found
at outskirts of large gas concentrations in star forming regions. A few
important physical effects are pointed out and discussed, including gas
dynamical friction, radiative efficiency of accretion flow, and
radiative feedback to ambient medium. ULXs could last for $\sim
10^{5-6}$ yr to emit a total energy of $\sim 10^{53}$ erg, which is
sufficient to power the ionized expanding nebulae found by
optical observations.
\end{abstract}

\section{Introduction}

Ultra-Luminous X-ray sources (ULXs, Makishima et al. 2000 and references
therein) are bright X-ray sources having luminosities greater than $\sim
3 \times 10^{39}\ergs$ found in off-nuclear regions of nearby galaxies.
The luminosity exceeds the Eddington limit of a $\sim 20 M_{\odot}$
black hole (BH) that is the maximum mass expected from normal stellar
evolutionary paths (Fryer \& Kalogera 2001), and their origin is now a
matter of hot debate. A few models have been proposed to explain the
high luminosity from ordinary stellar mass black holes, such as
geometrical beaming (King et al. 2001; King 2004) or super-Eddington
accretion (Begelman 2002). However, clear observational evidence for
these has not yet been found. On the other hand, the high luminosity may
also be explained by intermediate mass black holes (IMBHs) of $M_{BH}
\gtrsim 100 M_\odot$, without violating the Eddington limit. Recently
evidence for IMBHs has been accumulating for many ULXs. They show lower
temperature ($\sim$ 0.1 keV) blackbody components than that of the
Galactic black hole binaries, which is consistent with IMBHs being ULXs
(Miller et al. 2004). It is known that, for X-ray binaries and active
galactic nuclei (AGNs), the break frequency of the power density
spectrum of X-ray flux variability scales with the black hole mass, and
the break frequency of ULXs is indeed between those of black hole
binaries and AGNs, again indicating intermediate BH mass for ULXs
(Cropper et al. 2004).  It appears that at least a part of ULXs,
especially the most luminous ones, are IMBHs.  (See e.g., Colbert \&
Miller 2004; Mushotzky 2004 for recent reviews.)

However, the formation of IMBHs to become ULXs is a challenging problem.
One possibility is formation in dense stellar regions or stellar
clusters (Portegies Zwart \& McMillan 2002). However, though ULXs are
preferentially found in actively star forming regions, their position is
near but not coincident with such stellar clusters, disfavoring this
scenario (Zezas \& Fabbiano 2002).

On the other hand, it is theoretically expected that IMBHs form as
remnants of massive population III (Pop III) stars (e.g., Schneider et
al. 2002), and they may become ULXs if they accrete sufficient gas in
nearby galaxies (Madau \& Rees 2001). However, it is not easy to explain
the observed number of ULXs by this scenario. Islam, Taylor \& Silk
(2004) considered a possibility that free-floating IMBHs accrete from a
residual core of gas associated with the IMBHs from their formation
epoch. It is highly uncertain whether such gas core can
survive until present from the early universe, and this scenario cannot
explain the majority of ULXs that are associated with star forming
regions. Volonteri \& Perna (2005) recently performed a similar
calculation, and they found even more pessimistic prediction for the
observable ULX number. Krolik (2004) considered a possibility that
free-floating IMBHs become ULXs when they enter into dense molecular
clouds in star forming regions of galactic disks, so that the
association with star forming regions can be explained. Krolik
considered a hypothetical ``disk'' population of Pop III stars that form
within a galactic disk, from primordial gas that has somehow been
preserved until the disk formation. This is because the number of ULXs
would be too small if, as normally expected,  
Pop III remnants are distributed in a galactic
halo. It is an interesting possibility, but it is
difficult to quantitatively estimate how much amount of such gas
can survive during the formation of galactic disks.

The efficiency of Pop III star formation is highly uncertain, but there
are some observational hints. Independent groups (Wright \& Reese 2000;
Cambr\'esy et al. 2001; Matsumoto et al. 2004) reported detections of
the cosmic near-infrared background radiation (CNIB), which cannot be
explained by normal galaxy populations (Totani et al. 2001). The excess
is particularly large at $\sim 1.5 \mu$m, which is at $\sim 5$--$6
\sigma$ level according to the IRTS data of Matsumoto et al., though
there may still be large systematic uncertainty in the CNIB
measurements, such as zodiacal light subtraction. It has been shown that
efficient formation of Pop III stars before $z \sim 8$ can explan this
unaccounted excess of CNIB (Santos, Bromm, \& Kamionkowski 2002;
Salvaterra \& Ferrara 2003); the spectral peak of the excess at $\sim 1.5\mu$m
corresponds to the redshifted Lyman $\alpha$ line emission.

The required amount of Pop III stars to explain the CNIB is somewhat
extreme (see, e.g., Madau \& Silk 2005); about $\sim$ 10\% of the cosmic
baryons must be converted into Pop III stars. However, paucity of
theoretical understainding of Pop III stars does not allow to reject
this possiblity completely, and it is the only one realistic explanation
of the CNIB excess, except for the systematic uncertainty in the CNIB
measurements. Theoretical studies on Pop III star formation indicate
that they are likely more massive than normal stars, and most of Pop III
stars could be massive enough to collapse into IMBHs without any metal
ejection, meeting the metallicity constraints from quasar absorption
line systems. The consequence of this scenario is that there should be
IMBHs with $M_{BH} \sim 10^2$--$10^3 M_\odot$
in halos of nearby galaxies, whose mass fraction in the baryonic
matter is $f_{\rm III} \equiv \Omega_{\rm IMBH} / \Omega_B \sim 0.1$,
where $\Omega_B \sim 0.05$ is the cosmic baryon density in the standard
definition. This is consistent with observational constraints from
gravitational lensing experiments (Wambsganss 2002) and from stellar dynamics
such as heating of disk stars (Wasserman \& Salpeter 1994),
disruption of globular clusters (Murali et al. 2000), and disruption of
halo wide binaries (Yoo et al. 2004).

In this paper we show that, if this is the case, such IMBHs passing
through dense molecular clouds of star forming regions should have a
siginificant contribution to the observed ULX population, especially to
the highest luminosity ones. In \S \ref{section:order}, we give
an order-of-magnitude estimate of the expected ULX number.
We discuss two relevant physical processes in \S \ref{section:DF} (gas
dynamical friction) and \S \ref{section:radiative-efficiency}
(radiative efficiency of accretion disks). We will present
a more detailed calculation of the predicted ULX luminosity function
in \S \ref{section:LF}. Radiative feedback and the ULX duty cycle
will be discussed in \S \ref{section:feedback}. Finally some
discussions, conclusions, and predictions
will be presented in \S \ref{section:conclusion}.

\section{Number of ULX\lowercase{s}: An Order-of-Magnitude Analysis}

\label{section:order}

Detectability of isolated BHs accreting interstellar gas has been
studied in a number of publications (Fujita et al. 1998, Agol \&
Kamionkowski 2002 and references therein), and we follow these for the
basic formulations. As a typical galaxy, we choose parameters of our
Galaxy with a total halo mass of $M_{\rm halo} = 10^{12} M_\odot$ and
circular velocity of $V_c = 220\kms$.
Luminosity of a black hole with mass $M_{BH}$ expected from the
Bondi-Hoyle accretion in a gas cloud is given by
\begin{eqnarray}
L_X(n, \upsilon) &=& \dot m c^2 \eta
= \frac{4 \pi \eta c^2  G^2 M_{BH}^2
\rho}{(\upsilon ^2+
\sigma_{MC} ^2+c_s^2)^{3/2}} \nonumber \\
&=& 4 \times 10^{37} \eta_{-1} \ M_2^2 \ n_2 \ \tilde\upsilon_1^{-3} \ \ergs
,
\label{eq:mdt}
\end{eqnarray}
where $\dot m$ is the gas mass accretion rate, $\upsilon$ the IMBH
velocity relative to the gas cloud, and $\rho$, $\sigma_{MC}$, and $c_s$
are the density, the turbulent velocity, and the sound speed of a
molecular cloud, respectively. In the third equation, we used $M_2
\equiv M_{BH}/(100 M_\odot)$, $n_2 \equiv n/(100 \cc)$, and
$\tilde \upsilon _1 \equiv \tilde \upsilon /(10\kms)$, where $n = \rho /
(\mu m_p)$ is the hydrogen number density, $m_p$ the proton mass,
$\mu=0.13$ the mean molecular weight, and $\tilde \upsilon \equiv
(\upsilon^2 + \sigma_{MC}^2 + c_s^2)^{1/2}$. The parameter $\eta
\equiv 0.1 \eta_{-1}$ represents not only the radiative efficiency of
accreting gas but also includes the uncertainty in the Bondi-Hoyle
formula of accretion rate.  The turbulent velocity $\sigma_{MC}$ scales
with density as $\sigma_{MC} \simeq 3.7 n_2^{-0.35}\kms$ (Larson 1981).
The sound speed $c_s$ depends on the gas temperature, which is $c_s \sim
0.3 \kms$ for cold molecular gas ($T=$ 10 K) but $\sim 10 \kms$ for
ionized hot gas ($T \sim 10^4$ K).

The number of ULXs in the Galaxy, whose luminosity and ambient gas
density are $L_X$ and $n$, respectively, is then given as:
\begin{eqnarray}
\frac{\partial^2 N_{\rm ULX}}{\partial n \partial L_X} =
N_{\rm IMBH} \ f_{\rm disk} \ \frac{df_n(n)}{dn}
\frac{df_\upsilon(\upsilon)}{d\upsilon} \left| \frac{\partial\upsilon
(n, L_X)}{\partial L_X} \right| \ ,
\end{eqnarray}
where $N_{\rm IMBH} = f_{\rm III} (\Omega_B/\Omega_M) M_{\rm halo} /
M_{BH}$ is the total number of IMBHs in our Galactic halo,
the cosmological density parameter
$\Omega_M = 0.3$, $f_{\rm disk}$ the fraction of halo IMBHs passing
through the disk region, $df_n/dn$ the volume fraction of regions
in the Galactic
disk which is occupied by molecular clouds of density $n$
per unit cloud density, and
$df_\upsilon/d\upsilon$ the velocity distribution function of IMBHs. In
this equation, $\upsilon$ is a function of $n$ and $L_X$ as determined
by eq. (\ref{eq:mdt}).

Assuming the Navarro, Frenk, \& White (1997) density profile with
parameters given in Klypin, Zhao, \& Somerville (2002), we estimate the
dark matter mass included within 10 kpc radius is about 4\% of $M_{\rm
halo}$ \footnote{The infall of baryonic gas into the center of a dark
halo leads to adiabatic compression and a subsequent enhancement of dark
matter density including halo IMBHs, which will be discussed in \S
\ref{section:conclusion}.}.  Then $f_{\rm disk}$ can be estimated as
$\sim 0.04 \times (3/4) \ 2 H_g / (10\ \rm kpc) \sim 4.5 \times
10^{-4}$, where $H_g$ = 75 pc (Sanders, Solomon, \& Scoville 1984) is
the disk scale height of molecular gas.  The velocity of ULXs must be
much smaller than the dispersion of the Maxwell distribution,
$\sigma_{BH} = V_c/\sqrt{2} = $ 160 km/s, and hence $df_\upsilon /
d\upsilon \sim (2/\pi)^{1/2} \upsilon^2 / \sigma_{BH}^3$.

Following Agol \& Kamionkowski (2002), we model $f_n$ as $df_{n}/dn
\propto n^{-\beta}$ with $\beta = 2.8$, between $n_{\min}=10^2$ and
$n_{\max}=10^5\textrm{ cm}^{-3}$. Normalization is given as:
\begin{equation}
 \frac{df_n}{dn} = \frac{(\beta-2) \left\langle\Sigma_{MC}\right\rangle }{
  2 \mu m_p n_{\min}^2 H_g}
 \left(\frac{n}{n_{\min}}\right)^{-\beta} \ ,
\end{equation}
where $\left\langle \Sigma_{MC} \right\rangle$ 
is the mean surface mass density of molecular gas in the
Galactic disk, which is averaged for IMBHs passing through the disk as:
\begin{eqnarray}
\left\langle \Sigma_{MC} \right\rangle = \frac{\int 2 \pi r \ \Sigma_{MC}(r)
\ \rho_{DM} \ dr}{\int 2 \pi r \ \rho_{DM} \ dr} \ ,
\end{eqnarray}
where $r$ is the galactrocentric radius and $\rho_{DM}$ the
dark halo density. 
By using the data of Sanders et al. (1984) for $\Sigma_{MC}(r)$, we found
$\left\langle \Sigma_{MC} \right \rangle = 29 M_{\odot}\textrm{
pc}^{-2}$.

We cannot take all value of $n$ between $n_{\min}$ and $n_{\max}$,
because $\tilde \upsilon$ is limited as $\tilde \upsilon >
(\sigma_{MC}^2 + c_s^2)^{1/2}$ and $n$ must be larger than
\begin{eqnarray}
n_{\min}'
\equiv 2.5 \times 10^3 L_{39} \eta_{-1}^{-1} M_2^{-2}
\left[ \frac{ (\sigma_{MC}^2 +
c_s^2)^{1/2} }{(10 \kms)} \right]^3 \ \rm cm^{-3}
\end{eqnarray}
for $\upsilon(n, L_X)$ to have a
solution, where $L_{39} \equiv L_X / (10^{39}\ergs$). For $n$ greater
than $n_{\min}'$,
\begin{eqnarray}
\upsilon(n, L_X) \sim 3.4 \ \eta_{-1}^{1/3} M_2^{2/3}
n_2^{1/3} L_{39}^{-1/3} \kms \ ,
\end{eqnarray}
and hence we find
\begin{eqnarray}
\frac{ \partial^2 N_{\rm ULX} }{
\partial n \ \partial L_X } \ \propto \ n^{1-\beta} L_X^{-2}\ .
\end{eqnarray}
Therefore
integrating over $n$ between $\tilde n_{\min}$ and $n_{\max}$, where
$\tilde n_{\min} \equiv \max (n_{\min}, n_{\min}')$,
we find the expected total number of ULX as:
\begin{eqnarray}
&& N_{\rm ULX}(>L_X) \sim  \frac{dN}{dL_X} L_X \\ &\sim&
N_{\rm IMBH} f_{\rm disk} \frac{\tilde n_{\min}}{\beta - 2}
\frac{df_n(\tilde n_{\min})}{dn} f_\upsilon \left[
<\upsilon(\tilde n_{\min}, L_X)\right]
\\ &=& 1.1 \times 10^{-2} \ L_{39}^{-1} \ \eta_{-1} \ M_2 \left(\frac{\tilde
n_{\min}}{n_{\min}}\right)^{2-\beta}
\ .
\end{eqnarray}
If $(\sigma_{MC}^2 + c_s^2)^{1/2} \sim 10 \kms$, $n_{\min}'$ is larger
than $n_{\min}$ and the number is reduced to $N_{ULX} \sim 8.6 \times
10^{-4}$ for the $M_{BH}=100M_\odot$ case, but $n_{\min}' < n_{\min}$
and no reduction of $N_{\rm ULX}$ for $M_{BH}=10^3 M_\odot$.

There is no ULX in our Galaxy, and the number is known to be
proportional to star formation rate (SFR) of galaxies (Grimm, Gilfanov,
\& Sunyaev
2003).  Under a reasonable assumption that the amount of molecular gas
clouds in a galaxy is proportional to SFR of the galaxy, the number of
ULX should also scale with SFR\footnote{When the morphology of a
starburst galaxy is greatly different from that of our Galaxy, a simple
scaling of the ULX number with SFR may not be exactly correct.  However,
it is difficult to quantitatively discuss this effect and it is 
beyond the scope
of the paper.}. Then $f_n$ of a starburst galaxy like the Antennae is
$30$ times larger than that of our Galaxy. The number of ULXs in such a
galaxy is then estimated to be $\sim 3.3$ for $M_{BH} = 10^3 M_\odot$,
while in the Antennae $\sim10$ ULXs are observed above $10^{39}\ergs$ (Zezas
\& Fabbiano 2002).  This is an encouraging result; it should be noted
that the observed ULXs probably include ordinary X-ray binaries in the
luminosity range of $\sim$(1--3)$\times 10^{39}\ergs$.
In the following sections, we discuss two more physical processes; one
(gas dynamical friction) may enhance the ULX number and the other
(radiative efficiency of accretion) is important to reproduce the
observed luminosity function of X-ray sources.

\section{The effect of gas dynamical friction}\label{section:df}
\label{section:DF}

The expected number of ULXs is sensitively dependent on the
relative velocity $\upsilon$, since $L_X \propto \upsilon^{-3}$
and $f_\upsilon(<\upsilon)
\propto \upsilon^3$. Therefore, if there is a process
that reduces the speed of IMBHs in molecular clouds, it would
significantly enhance the expected ULX number. Dynamical friction
is a possible effect, and considering that the star formation efficiency
in molecular clouds is generally low (stellar to gas mass ratio
$M_{\rm star}/M_{MC} \sim
$ 0.01--0.1, Myers et al. 1986; Larson 1988), the dynamical friction
by gas will be more efficient than by stars. The gas dynamical friction
time scale is given by (Ostriker 1999)
\begin{eqnarray}
t_{\rm df} &\equiv& \frac{\upsilon}{ \dot \upsilon} = \frac{M_{BH} \upsilon}
{F_{\rm df}} = \frac{\upsilon^3}{4 \pi \xi G^2 M_{BH} \rho \ln
(r_{\max}/r_{\min})} \\
&=& 8.5 \times 10^8 \xi^{-1}
\upsilon_{1}^3 M_{2}^{-1} n_{2}^{-1}  \ \rm yr \ ,
\end{eqnarray}
where $F_{\rm df}$ is the friction force, $\upsilon_1 \equiv
\upsilon$/($10 \kms$), $r_{\max}$ and $r_{\min}$ are the sizes of
surrounding medium and the object receiving the force, respectively, and
$\xi$ is a parameter of order unity to take account of the uncertainty
in the formula. Here we took
$r_{\max} = $ 20 pc and $r_{\min} = 10^3$ km ($\sim$ Schwartzschild
radius\footnote{Gas within the accretion radius $r_A \sim G
M_{BH}/\upsilon^2 \sim 
10^{16} \rm \ cm$ is captured by the IMBH, and momentum loss by the 
accretion
leads to a similar formula that is different
only by numerical factors.})  for
the Coulomb logarithm, $\ln (r_{\max}/r_{\min}) \sim 27$.

This should be compared to the crossing time of molecular clouds,
\begin{eqnarray}
t_{\rm cross} \sim \frac{ L_{MC} }{ \upsilon }= 2.4 \times 10^6 n_{2}^{-0.9}
\upsilon_{1}^{-1} \ \rm yr \ ,
\end{eqnarray}
where the size of typical molecular clouds is given as $L_{MC}
\sim 24.7 n_{2}^{-0.9} \ \rm pc$ (Larson 1981).
The ratio
\begin{eqnarray}
\frac{ t_{\rm df} }{ t_{\rm cross} } = 350 \ \xi^{-1} \upsilon_1^4 \
n_2^{-0.1} M_2^{-1}
\end{eqnarray}
becomes of order unity for $M_{BH} \sim 10^3
M_\odot$ and $\upsilon \lesssim 5 \kms$. The dependence on $\upsilon$
and $n$ indicates that this is achieved only for objects having the
highest accretion rate, i.e., ULXs. Therefore this process could boost
up the ULX luminosity particularly at the brightest end of the ULX
luminosity function.

\section{Radiative Efficiency and Luminosity Function Slope}
\label{section:radiative-efficiency}

So far we assumed that the X-ray luminosity is proportional
to the Bondi accretion rate, i.e., a constant radiative efficiency,
$\eta$. 
However, in this case we have a serious
problem about the ULX luminosity function. Since $L_X \propto \dot m
\propto \upsilon^{-3}$, while $f_{\upsilon}(<\upsilon) \propto
\upsilon^3$, a robust prediction is that the luminosity function should
scale as $dN/dL \propto L^{-\alpha}$ with $\alpha = 2$. On the other
hand, the observed luminosity function of ULX in star forming galaxies
is smoothly
connected\footnote{This fact is often used to argue that ULXs are
physically an extension of normal HMXBs (e.g., Gilfanov 2004),
but the latest work on ULX
luminosity function indicates that ULXs with $L_X \gtrsim 10^{40}\ergs$
might be a different population from lower luminosity ones (Liu,
Bregman, \& Irwin 2005). In our scenario, the similarity between
the number of
IMBH-ULXs and extralopation of the HMXB luminosity function
must be a coincidence.}  to that of high
mass X-ray binaries (HMXBs), with a slope of $\alpha = 1.6$ in a wide
luminosity range of $L_X \sim 10^{36}$--$10^{39} \ergs$ (Grimm
et al. 2003).  Therefore, even if we successfully
explained the observed number of ULXs, it would overproduce the
observable number of low-luminosity X-ray sources by the same IMBH
population.

This problem can be solved if we take into account the change of
accretion mode at low accretion rate compared with the Eddington rate;
recent studies established that accretion disks change their state
from standard thin disks to radiatively inefficient accretion flows
(RIAFs), at luminosity below $\sim$ 10\% of the Eddington luminosity
(Kato, Fukue, \& Mineshige 1998; Narayan 2004).
In the RIAF regime, the luminosity scales roughly $L \propto
\dot m^2$, and hence the luminosity function slope of ULXs becomes
$\alpha = 1.5$, avoiding the overproduction problem of low luminosity
sources.

\section{Modeling the Luminosity Function}
\label{section:LF}

Here we calculate the ULX luminosity function by integrating numerically
over $n$, $\upsilon$, and the location of a BH in the Galaxy. The
calculation is based on the formulation of Agol \& Kamionkowski (2002),
but we also include the follwing two effects discussed above. (i) 
To make the luminosity proportional to $\dot m^2$ in the RIAF regime, the
radiative efficiency is reduced as $\eta = \dot m / (0.1 m_{\rm Edd})
\eta_{\rm Edd}$ when $\dot m < 0.1 \dot m_{\rm Edd}$, while
$\eta = \eta_{\rm Edd}$ for $\dot m > 0.1 \dot m_{\rm Edd}$, 
where $\dot m_{\rm Edd} \equiv
L_{\rm Edd}/(c^2 \eta_{\rm Edd})$, $L_{\rm Edd}$ the Eddington luminosity, 
and $\eta_{\rm Edd}$ is a constant
parameter to determine the efficiency around the Eddington accretion rate.
(ii) The relation $\upsilon (n, L_X)$ is
replaced by $\upsilon' (n, L_X)$ to take into account the dynamical
friction, where $\upsilon'$ and $\upsilon$ 
($\upsilon' > \upsilon$) are the velocity before and
after an IMBH passes through a molecular cloud having the size
$L_{MC}$ (see \S \ref{section:DF}),
whose relation is determined by solving the equation of motion with the
dynamical friction. Since the initial mass function (IMF) of Pop III
stars is highly uncertain, we simply calculate for a $\delta$-function
IMF with $M_{BH} = 10^2$ and $10^3 M_\odot$.

\begin{figure*}
\epsscale{1.0}
\plotone{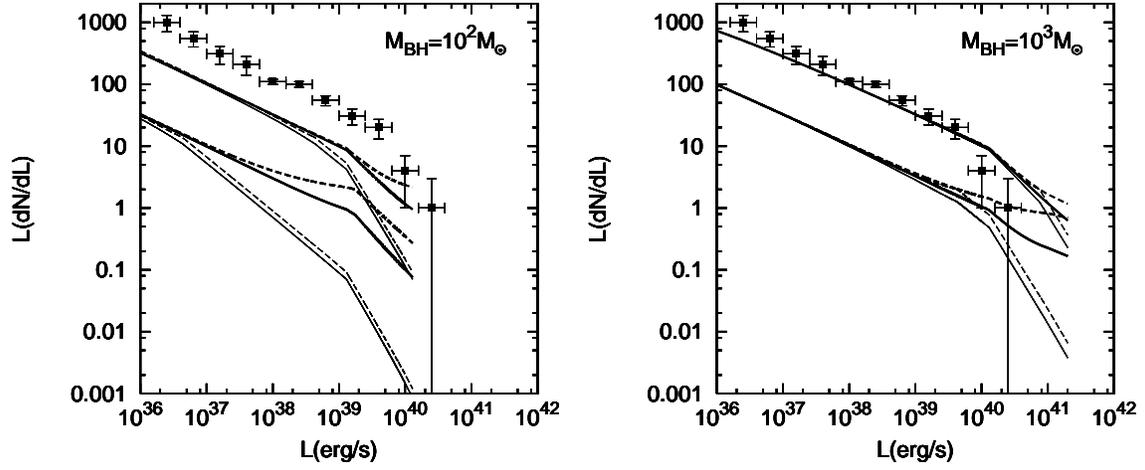}
\caption{{\em Left}: The X-ray luminosity function of
free-floating IMBHs ($M_{\rm BH}= $100 $M_\odot$) passing through
dense molecular clouds,
normalized to
SFR$_{>5}=50M_{\odot}~{\rm yr}^{-1}$. For each line marking,
there are two model curves: the upper one for $\eta_{\rm Edd} = 1$ and
the lower one for $\eta_{\rm Edd} = 0.1$. The sound speed is
set as $c_s = 0.3$ and 10 $\kms$ for
each thick and thin curve, respectively. The
dynamical friction parameter is set to be
$\xi=1$ and 10 for each solid and dashed curve, respectively.
The data points show the observed universal X-ray luminosity function
from Grimm et al. (2003).
{\em Right}: Same as the left panel, but for $M_{\rm BH}=10^3M_{\odot}$.}
\label{fig}
\end{figure*}

The results are shown in Fig.~\ref{fig}.  We scale up the luminosity
function of the Galaxy to that for a galaxy with SFR$_{>5} = 50
M_\odot~{\rm yr}^{-1}$, to compare with the observed data of the
universal luminosity function constructed by X-ray sources in different
galaxies and normalized by SFR, where SFR$_{>5}$ is SFR for stars
heavier than $5 M_\odot$ and SFR$_{>5} = 0.25 M_\odot~{\rm yr}^{-1}$ for
our Galaxy (Grimm et al. 2003).  Considering the uncertainties, we show
eight calculations with $\xi=1$ or $10$, $c_s=0.3\kms$ or $10\kms$, and
$\eta_{\rm Edd}=0.1$ or $1$. It should be noted that the parameter
$\eta_{\rm Edd}$ also includes the uncertainty in the Bondi-Hoyle
accretion formula. Still, the case of $\eta_{\rm Edd} =1$ is rather
unlikely, but we take this value as a maximally possible one. The choice
of $\xi = 10$ may also be extreme, but we took this value to show the
effect of dynamical friction clearly.
There are
two breaks in the model curves: one corresponding to the transition
luminosity between the two accretion modes (the standard thin disk or
RIAF) at $L_X = 0.1 L_{\rm Edd}$, and another corresponding to the
luminosity above which $n_{\min}'>n_{\min}$.  As expected, the dynamical
friction effect boosts up the luminosity function at the brightest
luminosity range.

These results show that at least a part of the brightest ULXs ($L_X
\gtrsim 10^{40}\ergs$) can be explained by our scenario, without
overproducing low luminosity sources in the range of $L_X \lesssim
10^{39}\ergs$, where normal HMXBs are dominant. 
(Therefore we do not have to explain all of the low luminosity sources.)
Especially, with $M_{BH}
= 10^3 M_\odot$, the number of the most luminous ULXs 
($L_X \gtrsim 10^{40}$ erg/s) can be explained
with a modest value of $\eta_{\rm Edd} =0.1$, favoring a larger IMBH
mass. The enhancement of IMBH density around galactic disks by 
baryon compression may further increase the number of ULXs, possibly
allowing a lower IMBH mass (see \S \ref{section:conclusion}).
If the radiative feedback and ionization of ambient gas are
siginificant, $c_s$ should be raised to $\sim 10\kms$ and the number of
ULX would be significantly reduced, especially in the case of $M_{BH} =
100 M_\odot$.  However, in the next section we will argue that this
effect is not necessarily strong.

\section{On the Radiative Feedback}
\label{section:feedback}

The ULX luminosity is sufficient to ionize the surrounding medium;
the Str\"omgren radius is
\begin{eqnarray}
r_S \sim 7.0 \times 10^{18} n_2^{-2/3} \epsilon_{-1}^{1/3}
L_{39}^{1/3} \ \rm cm \ ,
\end{eqnarray}
where $\epsilon$ is the fraction of ionizing luminosity above
13.6 eV compared with $L_X$, and $\epsilon_{-1} = \epsilon / 0.1$.
This is in most cases larger than the Bondi accretion radius,
\begin{eqnarray}
r_A \sim \frac{ G M_{BH} }{
\tilde\upsilon^2 } = 1.3 \times 10^{16} M_{2} \tilde\upsilon_{1}^{-2} \
\rm cm.
\end{eqnarray}
Therefore, when an ULX ionizes the surrounding medium, the Bondi
accretion rate should reduce, because of the increased $c_s$ and
expansion of gas driven by increased pressure. However, the gas that has
already accreted to form an accretion disk will serve as the energy
source of the ULX phenomenon for some time after the ionization. Based
on the interstellar density gradient and angular momentum conservation,
the interstellar gas captured by an IMBH should form a rotationally
supported disk at a radius $r_d \sim 7.5 \times 10^{13}
M_2^{5/3}\upsilon_1^{-10/3} \ \rm cm$ (Agol \& Kamionkowski 2002), which
is much smaller than the Bondi accretion radius, $r_A$.  Assuming that
the further accretion proceeds by $\alpha$-viscousty in the standard
thin disk, accretion rate for ULX-level luminosity is not achieved until
a sufficient amount of mass is accumulated and the surface mass density
$\Sigma$ at $r \sim r_d$ reaches that of the standard disk for the mass
accretion rate of ULXs, $\dot m$. It would take a time of $\sim 2 \pi
\Sigma r_d^2 / \dot m$. Then ULX activity will last for a time scale of
the disk mass consumption, $2 \pi \Sigma r_d^2 / \dot m$, even after the
ULX ionizes the surrounding medium.

Therefore both the time scale of turning on of ULXs and duration of the
ULX phenomenon are given by the accretion time scale of the standard
disk,
\begin{eqnarray}
t_{\rm sd} &\sim& \frac{ 2 \pi \Sigma r_d^2 }{ \dot m } \\
&\sim& 2.7 \times 10^5
\alpha_{-1}^{-4/5} M_2^{-4/9} L_{39}^{49/45} \eta_{-1}^{-49/45}
n_2^{-25/18} \ \rm yr \ .
\end{eqnarray}
(The outer region of the three regions in
the standard disk theory applies
at $r \sim r_d$.)  It is interesting that this time scale for an ULX
with $L_X \sim 10^{40}\ergs$ is similar to the crossing time scale $t_{\rm
cross} \sim 10^6$ yr of a $n \sim 10^2 \ \rm cm^{-3}$ cloud (\S
\ref{section:df}).  This means that the duty cycle as an ULX could
effectively be of order unity for an IMBH passing through a molecular
cloud, even if the radiative feedback is significant.

\section{Discussion and Conclusions}
\label{section:conclusion}

We conclude that the observed number of ULXs can be explained by halo
population IMBHs passing through dense molecular clouds, which were
produced as Pop III star remnants, if the Pop III star formation was
very efficient as recently indicated by the unaccounted excess of the
cosmic near-infrared background radiation.  Luminosity function was
calculated, and the ULXs of our scenario contribute to the observed
luminosity function especially at the largest luminosity range, without
overproducing sources in low-luminosity range where stellar-mass black
holes are the dominant contribution. The ULX phenomena could last for
$10^{5-6}$ yr and the duty cycle in a molecular cloud could be of order
unity.

We considered molecular clouds with a denstiy range of $10^2$--$10^5 \
\rm cm^{-3}$, but rather low-density clouds of $n \sim 10^2 \ \rm
cm^{-3}$ have the largest contribution to the expected ULX number.  It
is consistent with observations of ULX locations; many ULXs are found in
star forming regions but in outskirts of large concentrations of
gas (Zezas et al. 2002).

Large ionized, expanding optical nebulae with a size of $\gtrsim$ 100 pc
are often associated with ULXs, with estimated total expansion energy
of $\sim 10^{52}$ erg and age of $\sim 10^6$ yr (Pakull \& Mirioni 2002;
Mushotzky 2004). Our estimate of the ULX duration is consistent with
this age, and the
total energy emitted by an ULX with $L_X \sim 10^{40} \ergs$ could be as
large as $\sim 3 \times 10^{53}$ erg. A part of this energy may be
emitted in UV and soft X-ray bands to ionize and expand the ambient
medium, giving an explanation for the observed optical nebulae.

The estimated ULX duration is not very different from the Eddington time
scale, $t_{\rm Edd} = M_{BH} / (L_{\rm Edd}/\eta c^2) = 4.4 \times 10^7
\eta_{-1}$ yr.  Therefore, some ULXs may have a chance to accrete gas
for a time scale comparable to or longer than $t_{\rm Edd}$, by passing
through a large molecular cloud or many molecular clouds. Such IMBHs
will increase their mass considerably. In fact, the IMBH mass used in
our calculation is not necessarily that of the Pop III remnants.  The
mass function of IMBHs in the local universe may have been altered from
the original one at the formation epoch, by merging or gas accretion in
the hierarchical structure formation.  Such calculations have been
performed by Islam, Taylor, \& Silk (2004) and Volonteri \& Perna
(2005). According to these calculations, typical mass of free-floating
IMBHs may be increased up to $\sim 10^4 M_\odot$ depending on model
parameters, even if the initial mass of Pop III remnant BHs is less than
1000 $M_\odot$. The very large BH mass suggested for the extremely
bright ULX in M82 (Matsumoto et al. 2001; Kaaret et al. 2001) may be
achieved by such processes.

We used the popular NFW density profile for the Galactic halo, but the
dark matter density could be significantly altered and increased around
the region of the Galactic disk, by adiabatic compression during the
infall of baryonic gas to form stars. The number of ULXs would then be
increased, because more IMBHs exist around the Galactic disk and the
weighted mean surface density of molecular gas is also increased.
According to the model by Klypin et al. (2002), we estimate the boost
factor of the ULX number as $\sim 3$. On the other hand, the same
authors also shown that a modest exchange of angular momentum between
baryonic and dark matter (through e.g., a rotating bar-like bulge)
may result in almost no enhancement, or even decrease of the dark matter
density. This should be taken into consideration as an uncertainty
in the prediction made here, though the main conclusions are unlikely
to be changed.

Finally we give a list of the predictions for ULXs produced by the
scenario proposed here. Since ULXs
are isolated IMBHs, optical counterparts (donor stars) will not be found.
Future observations should confirm the trend that ULXs
are preferentially found in star forming regions, but at outskirts of
dense molecular clouds where the density is about $\sim$ 100 $\rm
cm^{-3}$. Expanding nebulae with kinetic energy of $\sim 10^{52}$ erg
should often be associated, {\it without} evidence for a past hypernova or
multiple supernovae. The predicted slope of the IMBH luminosity function
is $\alpha = 1.5$ at low luminosity range, which is similar to that of
HMXBs. Therefore, in a wide range of luminosity,     
we expect that a small fraction of low-luminosity X-ray
sources are the same population with IMBH-ULXs, showing
RIAF spectra without companion stars.
It is interesting to search for such a population
in the Galaxy or nearby galaxies.  The excess of CNIB must be confirmed
with greater significance. The high density of free-floating IMBHs in
galactic haloes may be detected by future gravitational lensing
experiments (e.g., Totani 2003).

We would like to thank the anonymous referee for useful comments.
This work has partially been supported by the Grant-in-Aid for the 21st
Century COE ``Center for Diversity and Universality in Physics'' from the
Ministry of Education, Culture, Sports, Science and Technology (MEXT) of
Japan.

\clearpage

\end{document}